\documentclass[prb,aps,twocolumn,superscriptaddress,showpacs]{revtex4-1}

\usepackage{graphicx}
\usepackage{epstopdf}

\begin{document}

\title{Neutron scattering measurements of $dd$ and spin-orbit excitations below the Mott-Hubbard gap in CoO}

\author{R. A. Cowley}
\affiliation{Department of Physics, Clarendon Laboratory, Parks Road, Oxford, OX1 3PU, UK}
\author{W. J. L. Buyers}
\affiliation{National Research Council, Canadian Neutron Beam Centre, Chalk River, ON K0J 1JO, Canada}
\affiliation{Canadian Institute for Advanced Research, Toronto, ON M5G 1Z8, Canada}
\author{C. Stock}
\affiliation{School of Physics and Astronomy, University of Edinburgh, Edinburgh EH9 3JZ, UK}
\author{Z. Yamani}
\affiliation{National Research Council, Canadian Neutron Beam Centre, Chalk River, ON K0J 1JO, Canada}
\author{C. Frost}
\affiliation{ISIS Facility, Rutherford Appleton Labs, Chilton Didcot, OX11 0QX, UK}
\author{J. W. Taylor}
\affiliation{ISIS Facility, Rutherford Appleton Labs, Chilton Didcot, OX11 0QX, UK}
\author{D. Prabhakaran}
\affiliation{Department of Physics, Clarendon Laboratory, Parks Road, Oxford, OX1 3PU, UK}
\date{\today}

\begin{abstract}

Neutron scattering is used to investigate the single-ion spin and orbital excitations below the Mott-Hubbard gap in CoO.  Three excitations are reported at 0.870$\pm$0.009 eV, 1.84$\pm$0.03, and 2.30$\pm$0.15 eV.  These were parameterized within a weak crystal field scheme with an intra-orbital exchange of $J(dd)$=1.3 $\pm$ 0.2 eV and a crystal field splitting $10Dq$=0.94 $\pm$ 0.10 eV.  A reduced spin-orbit coupling of $\lambda$=-0.016$\pm$ 0.003 eV is derived from dilute samples of Mg$_{0.97}$Co$_{0.03}$O, measured to remove complications due to spin exchange and structural distortion parameters which split the cubic phase degeneracy of the orbital excitations complicating the inelastic spectrum.  The 1.84 eV, while reported using resonant x-ray and optical techniques, was absent or weak for non resonant x-ray experiments and overlaps with the expected position of a $^{4}A_{2}$ level.  This transition is absent in the dipolar approximation but expected to have a finite quadrupolar matrix element that can be observed with neutron scattering techniques at larger momentum transfers.  Our results agree with a crystal field analysis (in terms of Racah parameters and Tanabe-Sugano diagrams) and with previous calculations performed using local-density band theory for Mott insulating transition metal oxides.  The results also demonstrate the use of neutron scattering for measuring dipole forbidden transitions in transition metal oxide systems.

\end{abstract}


\maketitle

\section{Introduction}

The properties of $3d$ transition metal systems have been understood in terms of the Mott-Hubbard model where charge fluctuations are suppressed owing to a large electronic Hubbard gap.  This model forms the basis of our understanding of the electronic properties in high temperature superconductors and also the  interactions in multiferroic materials.~\cite{Lee06:78}  Optical techniques have been used to verify these models, but are limited by selection rules requiring $\Delta l=\pm 1$ ($l$ is the orbital angular momentum) and have supplied restrictive information about $3d$ transition materials as $dd$ transitions can involve a change in the spin or no change in $l$.~\cite{Lorenzana95:74,Loudon:book}  Such techniques are also sensitive to a variety of cross sections, including strong multiple phonon scattering, which often mask the subtle $dd$ transitions.~\cite{Saitoh01:410,Gruninger02:418} Here we use neutron scattering to measure the magnetic excitations within the Mott-Hubbard gap of CoO to illustrate that CoO is a classical Mott-Hubbard insulator where the excitations are based upon crystal field excitations within the large Mott gap.  

CoO has a high temperature cubic rock salt structure and below a temperature of 295 K, the magnetic structure becomes antiferromagnetic, accompanied by a structural transition to a predominately tetragonal unit cell with $1-c/a=0.011$ at 10 K.~\cite{Greenwald53:6,Jauch01:64}  The electronic structure has both spin and orbital moments making CoO unusual when compared with other transition metal materials such as nickelates or cuprates.~\cite{Neubeck01:62,Kernavanois03:15,Tomiyasu04:70} The antiferromagnetic structure is uncertain with both collinear and non-collinear order describing the neutron diffraction data equally well.~\cite{Roth58:110,Laar65:138,Ressouche06:385}  Inelastic neutron scattering has shown the existence of low lying magnetic branches corresponding to coupled spin and orbital moments.~\cite{Sakurai68:167,Yamani08:403,Yamani10:88}  However, no complete theory including both spin and orbital moments with nearest and next nearest exchange terms, the magnetic structure, spin-orbit splitting, and crystal field parameters have been developed.  

Part of the failure to understand the low-energy properties of CoO is the ambiguity behind the electronic ground state of the Co$^{2+}$ ion and, in particular,  the role of spin-orbit coupling which is difficult to extract and disentangle in the presence of strong exchange terms in the Hamiltonian.  Original calculations based upon LDA+U resulted in energy scales too large to explain the observed spectrum obtained from non-resonant inelastic x-ray scattering.~\cite{Anisimov91:44,Larson07:99}  Further calculations showed that the observed energy scales involved orbital transitions and not the large Mott Hubbard gap $U\sim5$ eV.~\cite{Haverkort07:99}   However, other studies using ARPES suggested the possibility of charge transfer excitations where oxygen states play a key role in the band structure leading to the proposal of an additional ``Zhang-Rice" band.~\cite{Yin08:100}  Neutron scattering has a complementary cross section which is sensitive to both orbital and spin transitions with the selection rules determined by matrix elements of the operator $\vec{M}=\vec{L}+2\vec{S}$.  It is also well suited for low-energy transitions allowing the direct observation of the spin-orbit energy scale.

\begin{figure}[t]
\includegraphics[width=8.5cm] {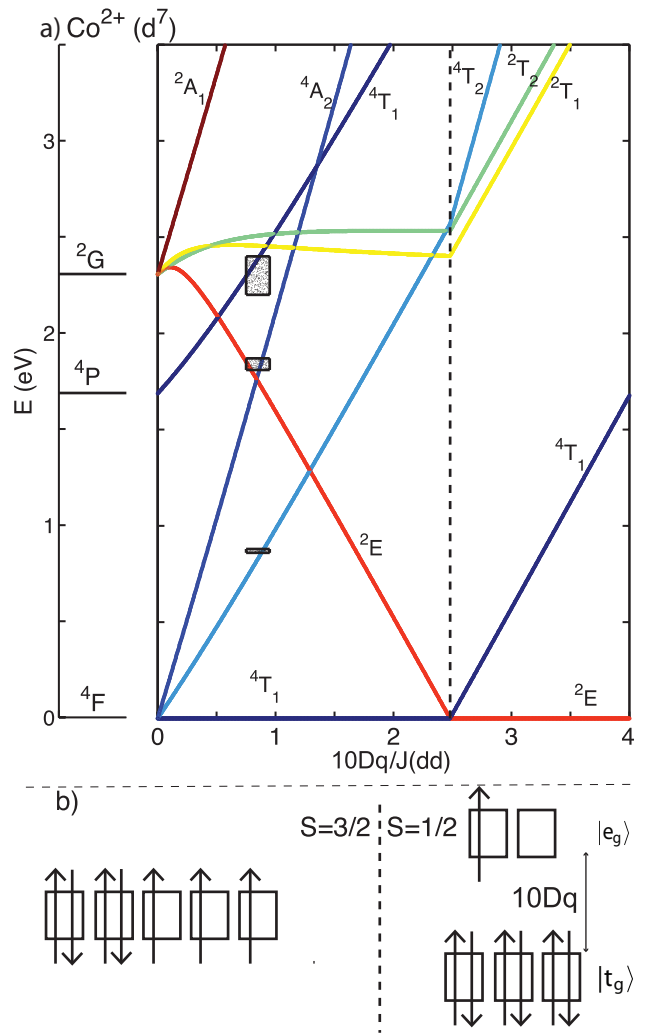}
\caption{\label{summary} $a)$ the energy values (Tanabe-Sugano diagram) for Co$^{2+}$ as a function of $10Dq/J(dd)$.  The shaded rectangles represent the excitations measured with neutrons at T=300 K using the MAPS spectrometer.  The vertical extent is the experimental error.  The states listed on the far left ($^{4}$F, $^{4}$P, and $^{2}$G) are the free ion solutions (Ref. \onlinecite{Racah42:62}). $b)$ Schematic illustration of the ground state for the extreme values of the crystal field splitting $10Dq / J(dd)$ (where $10Dq$ is the crystal field splitting and $J(dd)$ is the intra-orbital exchange defined in Ref. \onlinecite{Marel88:37}) where $S={3 \over 2}$ (when $10Dq/J(dd)$ is small),  and $S={1 \over 2}$ , (when $10Dq$ is large).}
\end{figure}

This paper reports a neutron scattering study of the single-ion properties of CoO with the ultimate objective of understanding the relative energy scales of the crystal field imposed by the neighboring oxygen ions and the spin-orbit coupling.  We have used high-energy neutron spectroscopy to probe the $\sim$ 1 eV single ion excitations characterizing the crystal field component and energy scale.   Dilute samples and lower energy neutron scattering was used to measure the much smaller energy of the spin-orbit coupling term.  The use of dilute samples removes the complex exchange terms which give a strong momentum dependence and a complex energy spectra.   We do not study the exchange terms in this paper which require a detailed measurement of the momentum dependence of the low-energy magnetic fluctuations with neutrons.   The key results of this paper are summarized in the Tanabe-Sugano diagram illustrated in Fig. \ref{summary}.  We will show, using neutron inelastic scattering data, that CoO resides in the weak-intermediate crystal field limit and discuss the transitions observed and predicted.  These will be compared with the various theories describing the competing crystal field and electron-electron repulsion energies.

The paper is divided into 5 sections including this first introductory component.  In the second section,  the experimental details are outlined including sample preparation, instrument configurations, and also background subtraction and data analysis.  The third section provides an outline of the relevant crystal field theory and also compares the Racah parameters derived from Tanabe-Sugano diagrams to more recently developed LDA+U calculations.  The fourth section describes the experimental results aimed at providing a complete experimental understanding of the single-ion properties in CoO.

\section{Experimental Details}

This section discusses the samples including preparation and thermodynamic characterization for the neutron experiments.  A discussion of the methods used to subtract the background are also provided.

\subsection{Samples and preparation}

The 10 g single crystal of CoO  was grown using the floating zone technique at Oxford University. The sample was prepared by annealing high purity Co$_{3}$O$_{4}$ ($>$99.99\%) under an atmosphere of high purity argon at a temperature of 1200$^{\circ}$C for 36 hours. The powder was reground at various intermediate times. Powder diffraction was used to establish the absence of any measurable ($\sim$ 1-2 \%) secondary Co$_{3}$O$_{4}$ phase. The CoO powder was then compressed into rods with a hydraulic press and the rods annealed at 1275$^{\circ}$C for 24 hours under an Ar atmosphere in a horizontal annealing furnace. Single crystals, 8 mm diameter and 100 mm in length, were then grown in a four-mirror optical floating-zone furnace. The growth was at a speed of 2-4 mm/hr with the feed and seed rods counter rotating at 35 rpm in an Ar atmosphere. Initially, a polycrystalline seed rod was used but on subsequent runs was replaced by a single crystal seed from earlier runs. 

\begin{figure}[t]
\includegraphics[width=7.5cm] {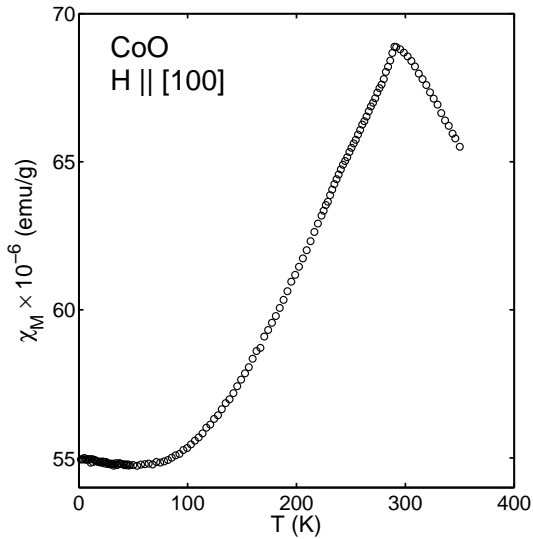}
\caption{\label{chi} The magnetic susceptibility measured on a 0.28 g sample in an applied magnetic field of 100 G along the [100] axis.  No measurable Co$_{3}$O$_{4}$ impurity phase is present as evidenced by the lack of any anomaly at low temperatures $\sim$ 30 K.}
\end{figure}

The crystal was cut to a length of 50 mm and used in the experiments described in this paper and for the lower energy studies described earlier.~\cite{Yamani08:403,Yamani10:88}  The nature of the surface of the CoO crystal was examined with optical and scanning electron microscopy and the results showed there was only one growth domain. X-ray scattering measurements showed that the surface had a mosaic spread of about 0.1$^{\circ}$ and no evidence was found at T = 295 K for multiple magnetic or strain domains. Finally, a small part of the sample was studied with a SQUID magnetometer and the results for the magnetic susceptibility are shown in Fig. \ref{chi}. The data agree well with previously published measurements.~\cite{Singer56:104}  The phase transition at 295 K is the antiferromagnetic transition of CoO and the absence of any discontinuities below 20 K shows that there is no significant amount of Co$_{3}$O$_{4}$ ($<$ 0.5\%) impurity.   

We also used a powdered sample of Co$^{2+}$ and MgO.  The samplewas prepared by mixing the appropriate amounts of Co$_{3}$O$_{4}$ and MgO in a ball mill for 5 hours.  The powder was annealed at 1200$^{\circ}$C in a flowing argon atmosphere for 36 hr with intermediate grinding. The powder was then examined by x-ray diffraction and no observable ($\sim$ 2\%) secondary phase was present.   The 45.8 g of material was placed in an airtight Al can under helium. The pure CoO powder used in the experiment had a mass of 15.7 g, and was also sealed in a similar manner.  The pure MgO powder was annealed at 1250 K for 12 hours and placed in another air-tight Al can. The MgO mass was 32.5 g.  These precautions involving high temperature annealing and sealing of the samples were taken because the fine powder was susceptible to water absorption producing spurious high-energy excitations related to hydrogen motion.

\subsection{Neutron scattering details}

The experiments mostly described here were performed using the MAPS and MARI direct geometry (fixed initial energy) chopper spectrometers located at the ISIS spallation neutron source.  The high-energy measurements were made on the MAPS spectrometer with fixed incident energies of E$_{i}$=2 eV and 3.5 eV.  The $t_{0}$ chopper spun at 100 Hz in parallel with the ``sloppy" Fermi chopper which was operated at a frequency of 600 Hz.   Lower incident energies were used to search for other crystal field excitations and to investigate the spin-orbit splitting in powder samples.  The $t_{0}$ chopper was spun at the nominal repetition rate of 50 Hz for these experiments with incident energies of E$_{i}$=750 meV, 250 meV, and 85 meV.  The Fermi chopper was spun at  600 Hz for all incident energies.   A bottom loading displex cooled the sample to temperatures as low as 15 K.

For lower energies, the MARI direct geometry spectrometer was utilized to study the powder samples of MgO and Mg$_{0.97}$Co$_{0.03}$O.  The $t_{0}$ chopper was operated at 50 Hz in parallel with a Gd chopper operating at 300 Hz with an incident energy of 85 meV.   A thick disk chopper with a frequency of 50 Hz reduced the background from high-energy neutrons.   A top loading displex cooled the sample to 5 K. Searches for low-energy crystal fields below the gapped fluctuations were performed on powders of CoO with E$_{i}$=100 and 10 meV with the Gd Fermi chopper spun at f=350 and 250 Hz respectively.

The lowest energy experiments were performed with the IRIS indirect geometry spectrometer at ISIS.  The final energy was fixed at E$_{f}$=1.84 meV by means of a cooled PG002 analyzer crystals in near backscattering geometry.  Through the use of three different time windows, the energy transfers of the neutron scattering could be measured up to $\sim$ 2 meV.  A top loading displex was used to cool the sample to 11 K.

For the high-energy experiments on MAPS, the single crystal CoO sample was aligned such that reflections of the form (H,H,L) lay within the horizontal scattering plane.  The magnetic signal was extracted from constant energy cuts as discussed in the next section.   The sample was nominally aligned with the [001] axis parallel to $\vec{k}_{i}$.  No angular dependence was observed when the sample was rotated as discussed below in the experimental results section and therefore all the angles were averaged to improve statistics.  

\subsection{Background subtraction and definition of Q$_{\parallel}$ and Q$_{\perp}$}

\begin{figure}[t]
\includegraphics[width=8.5cm] {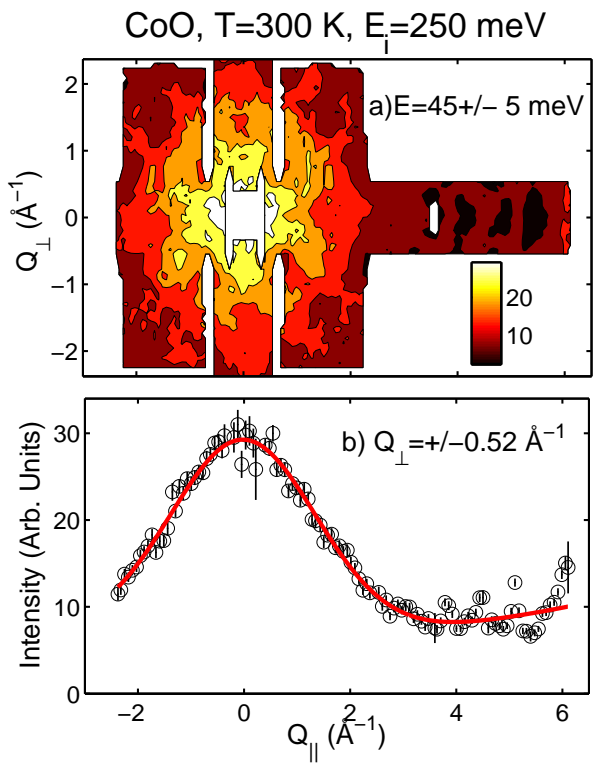}
\caption{\label{Ei_250meV} $a)$ constant energy slice through the paramagnetic scattering of CoO at E=45 $\pm$ 5 meV.  $b)$ illustrates a cut along the horizontal detector bank integrating over $\pm$ 0.52 \AA$^{-1}$ in the out of plane detectors.  The magnetic scattering is evidenced by the intensity centered on $Q$=0 that shows a decrease of intensity with increasing momentum transfer characteristic of magnetic scattering.}
\end{figure}

Background originating from the tails of the elastic line and multiple phonon scattering result in a steeply decreasing intensity curve as illustrated through out this text and in particular highlighted in Figs. \ref{Ei_2eV} $a)$ and \ref{Ei_3p5eV} $a)$ of the experimental section presented below.  To remove this background and extract the purely magnetic signal, we relied on the fact that the magnetic contribution to the scattering decreases with wavevector transfer while any background contribution will remain constant or increase with momentum transfer.  Assuming that the background at large scattering angles is dominated by phonons, the scattering was fitted at each energy transfer to the form,

\begin{eqnarray}
I(Q)=Ae^{-(Q/a)^{2}}+B+CQ^{2},
\label{background}
\end{eqnarray}

\noindent where $A$ characterizes the strength of the magnetic scattering, $a$ is proportional to the full width of the scattering in momentum, $B$ is a constant background, and $C$ characterizes the phonon contribution from both single and multi-phonon scattering.  As seen in Fig. \ref{Ei_250meV} and the representative scans in Figs. \ref{Ei_2eV}  and  \ref{Ei_3p5eV}, the fit provides a good description of the wave vector dependence of the scattering. 

The procedure is shown in Fig. \ref{Ei_250meV} which illustrates a constant energy slice and cut at E=45 meV in the paramagnetic phase of CoO  projected onto a plane with axes Q$_{||}$ and Q$_{\perp}$ where Q$_{||}$ is aligned along  [1, 1, 0]  and Q$_{\perp}$ is aligned along [1, $\overline{1}$, 0].  Q$_{||}$ is defined to lie in the scattering plane and therefore points in the direction of the high-scattering angle bank on MAPS which extends out to $\sim$ 60$^{\circ}$. Likewise, Q$_{\perp}$ is directed along the vertical direction.   The total momentum transfer $Q$ in terms of Q$_{||}$ for the in plane detectors (where the average of Q$_{\perp}\equiv$0) is equal to 

\begin{eqnarray}
Q^{2}=k_{i}^2+k_{f}^{2}-2k_{i}k_{f}\sqrt{1-\left({Q_{||}\over k_{f}}\right)^2}.
\label{momentum}
\end{eqnarray}

\noindent   The data shown in Figs. \ref{Ei_2eV}  and  \ref{Ei_3p5eV}, for E$_{i}$=2.0 eV and 3.5 eV used an integration in Q$_{\perp}$ of $\pm$ 0.52 \AA$^{-1}$ and  $\pm$ 2.1 \AA$^{-1}$, respectively. 

The use of this coordinate system is displayed in Fig. \ref{Ei_250meV} $b)$ which shows a cut  along Q$_{||}$ integrating over the vertical direction of Q$_{\perp}$ by $\pm$ 0.52 \AA$^{-1}$.  The cut shows a peak centered at Q$_{||}$=0 with significant width demonstrating it is of magnetic origin.  The solid red curve is a fit to Eqn. \ref{background} and provides a good description of the momentum dependence of the data.

For the incident energies discussed in Figs. \ref{Ei_2eV}  and  \ref{Ei_3p5eV}, we have performed similar constant energy scans and fits to remove the background.  The background corrected plots in Figs. \ref{Ei_2eV} and \ref{Ei_3p5eV} illustrate the integrated magnetic intensity derived from such fits.  

\section{Crystal field theory}

In this section of the paper, we outline the relevant aspects of the crystal field theory required to understand the single-ion excitations in CoO.   There are two important energy scales - the cubic crystal field and the spin orbit coupling.  The Hamiltonian can therefore be written as,

\begin{eqnarray}
H=H_{cef}+H_{so}.
\label{cubic_cef}
\end{eqnarray}

\noindent CoO is cubic at high temperatures and undergoes a small ($1-c/a\sim 0.01$ ) structural transition to a tetragonal unit cell below 295 K.  For the discussion here, which describes data in the cubic phase at T=300 K, we will not consider the effect of the tetragonal distortion on the Hamiltonian.

We discuss the crystal field theory of Co$^{2+}$ in CoO in terms of the methods outlined in Ref. \onlinecite{Abragam:book}. There are two possible approaches to the crystal field problem; the first is to assume that the effect of the crystalline field from the neighboring ions is small compared with the differences between the free ion energy levels; the second approach is to assume that the crystal field is large compared with the splitting of the atomic levels. We describe both of these approaches known as the weak and strong crystal field theory respectively.  As found from our experiments, cobalt oxide is an intermediate-weak case for which the crystal field is comparable to the energy differences between the atomic levels, while remaining in a high-spin state.

We first review the free ion solution for an isolated Co$^{2+}$ and write the solution in terms of Racah parameters ($A,B,C$), Slater integrals ($F^{i}$), and the Hubbard Model ($J(dd),C(dd),U$). 

\subsection{Free ion solution}

The cobalt ion, Co$^{2+}$, has the electronic structure given by $1s^{2}2s^{2}2p^{6}3s^{2}3p^{6}3d^{7}$. This structure consists of either filled or empty electronic shells apart from the $3d$ shell which can accommodate 10 electrons but which is occupied by only 7.   The energy eigenvalues of the isolated $d^{7}$ are degenerate in the absence of electron-electron interactions, however are split in the presence of these Coulomb interactions.  Historically, there have been several ways to parameterize this interaction with the original theory by Slater followed by a different parameterization of Racah.  More recently, this theory has been cast in terms of the Hubbard model.  We outline the three different parameterizations below for the free ion case.

The free ion solution, in the absence of a crystalline electric field that is present in a solid, for $3d^{7}$ can be written in terms of the Racah parameters $(A,B,C)$ (Ref. \onlinecite{Racah42:62}).  The thee lowest energy states then have the energies,

\begin{eqnarray}
\label{solutions}
E(^{4}P)=3A  \\
E(^{4}F)=3A-15B \nonumber\\
E(^{2}G)=3A-11B+3C \nonumber
\end{eqnarray} 

\noindent These parameters are a different way of writing the solution that was originally derived in terms of Slater integrals.  The Racah parameters can be recast in terms of the Slater integrals as,

\begin{eqnarray}
A=F_{0}-49F_{4} \\
B=F_{2}-5F_{4} \nonumber\\
C=35 F_{4}. \nonumber
\label{ABC}
\end{eqnarray}

\noindent However, another definition for the Slater terms $F^{0,2,4}$ (Ref. \onlinecite{Ballhausen:book}) is also sometimes used and is related to the Slater terms $F_{0,2,4}$ used above by the following relations,

\begin{eqnarray}
F_{0}=F^{0}\\
F_{2}={1\over 49}F^{2} \nonumber \\
F_{4}={1\over 448}F^{4}. \nonumber
\label{ABC}
\end{eqnarray}

\noindent The parameters above allow two different means of describing the free ion energy values for $3d^{7}$ Co$^{2+}$ and indeed other transition metal ions.  Physically, the terms $B$ and $C$ represent the effect of the interactions between the electrons in the $3d$ orbitals.  Experimentally, it is found that the ratio $\gamma\equiv C/B$ is close to a constant equal $\sim$ 4.6 for the $3d$ ions and therefore the Racah parameter $B$ is taken as a means of solely quantifying electron correlations.  

Based on this observation, it is convenient to introduce a single parameter that captures the effect of electron correlations that characterizes the intra-orbital exchange.   In terms of the Slater integrals listed above (which can also be written in terms of Racah parameters) an intra-orbital exchange can be defined which describes the attraction of parallel spins in different $d$ orbitals (Ref. \onlinecite{Marel88:37}).

\begin{eqnarray}
J(dd)={1\over 14} \left(F^{2}+F^{4}\right)
\label{Jdd}
\end{eqnarray}

\noindent  Using the above expressions, $J(dd)$ can be written in terms of either Slater ($F^{i}$) integrals or Racah ($B,C$) parameters.  Ref. \onlinecite{Marel88:37} also defines an additional quantity $C(dd)$ (not to be confused with the Racah parameter $C$) and it is defined as follows.

\begin{eqnarray}
C(dd)={1\over 14} \left({9\over7}F^{2}-{5\over7}F^{4}\right).
\label{Cdd}
\end{eqnarray}

\noindent This quantity physically represents an angular part of the multiplet splitting.  A third quantity known as the Hubbard $U$ is also defined and can be written in terms of the integral $F^{0}$ and $J(dd)$ as outlined in Ref. \onlinecite{Marel88:37}.  This parametrization is particularly useful as the Hund's rule ground state energy can be written as $F^{0}-J(dd)+C(dd)$ for $d^{7}$.  Noting that $C(dd)\sim 0.5 J(dd)$ (Ref. \onlinecite{Marel88:37}) this reduces to $F^{0}-0.5J(dd)$ illustrating the fact that a larger $J(dd)$ favors a Hund's rule ground state - a result demonstrated schematically in our Tanabe-Sugano diagram (Fig. \ref{summary}).  

The above summary demonstrates three ways of presenting the solutions for the free ion levels of $3d^{7}$.  The first is in terms of Racah parameters ($A,B,C$), the second in terms of Slater terms ($F^{0},F^{2},F^{4}$), and the third in terms of parameters $J(dd), C(dd),U$.  It can be seen that the Racah parameter $A$ and the Hubbard $U$ cannot be directly probed in our experiment that is only sensitive to transitions between energy levels.  This is obvious from the solutions written in Eqn. \ref{solutions} which all depend on $3A$.  Therefore, our neutron scattering experiments are only able to determine Racah parameters $B,C$ and also the more recent parameterization values of $J(dd),C(dd)$.

\subsection{Cubic crystalline electric field}

\subsubsection{Weak crystal field theory}

In weak crystal fields the ground state of the electronic wave functions for the 3d electrons is given by Hund's rules as for the atomic ground states. The first rule states that the electrons in any partially occupied shell should be aligned so that the total spin $S$ is a maximum, consistent with the Pauli principle. This implies that five of the electrons will have parallel spins with the two additional electrons having oppositely aligned spins giving a total spin of S = $({5\over2} - 1) = 3/2$ as illustrated in the left side of Fig. \ref{summary} $b)$. The second of Hund's rules states that the total orbital angular momentum, $L$, will be maximized subject to the constraints caused by application of the first of Hund's rules. This implies that the first 5 electrons half-fill the 3d orbitals and that the orbital angular momentum of these electrons is zero. The remaining two electrons have the maximum total angular momentum so that $L = (2+1) = 3$. In the spectroscopic notation the ground state of the isolated Co$^{2+}$ ion the ground state is then $^{4}F$ (not to be confused with the Slater $F$ discussed above).  The higher energy free ion states are illustrated on the left side of Fig. \ref{summary} $a)$ with the energy scale set by the Racah parameters $B$ and $C$.

Taking the $^{4}F$ ground state for Co$^{2+}$, we now consider the effect of a crystalline electric field resulting from the symmetry imposed by the lattice of CoO.  In the absence of mixing between the ground  $^{4}F$ state and the excited levels (such as the $^{4}P$), the cubic crystal field is given by the following Hamiltonian cast in terms of Stevens operators ($O_{4}^{0}$ and $O_{4}^{4}$) and numerical coefficients ($B_{4}$) (Refs. \onlinecite{Hutchings65:16,Walter84:45}),

\begin{eqnarray}
H_{cef,cub}=B_{4}(O_{4}^{0}+5O_{4}^{4}).
\label{cubic_cef}
\end{eqnarray}

\noindent In the high temperature cubic phase of CoO, the octahedral cubic symmetry splits the 7 orbital states of the lowest energy Hund's rule $^{4}F$ state into two orbital triplets ($^{4}T_{1}$  and  $^{4}T_{2}$) and one singlet ($^{4}A_{2}$) as illustrated in the Tanabe-Sugano diagram plotted in Fig. 1 $a)$.  For small values of $B_{4}$, the energy splitting between the levels are $\Delta E(^{4}T_{1} \rightarrow ^{4}T_{2})=480 B_{4}\equiv8 Dq$ and $\Delta E(^{4}T_{2}\rightarrow^{4}A_{2})=600 B_{4}\equiv10 Dq$.  

\subsubsection{Strong crystal field theory}

An alternative approach, to that used in the preceding section and Fig. 1 $a)$, is to assume that the energy splittings caused by the crystal field are much larger than the difference in energy between the free-ion states.  The analysis is similar to that described above, but instead of taking the free ion states as our basis we take the strong crystal field basis where the 3$d$ orbital states are split into triplet $|t\rangle$ and doublet $|e\rangle$ states with an energy difference defined by $\Delta$.  In an octahedral field, the $|t\rangle$ states have the lowest energy ($-{2\over5} \Delta$) while the $|e\rangle$ state is the highest (${3\over5} \Delta$).

The ground state electronic configuration can be derived using Hund's rules in the context of a large energy gap between the $|t\rangle$ and $|e\rangle$ states.  In such a scenario, it becomes energetically favorable to fill up the $|t\rangle$ states before populating the $|e\rangle$ levels.  This gives rise to a ground state illustrated in the right hand portion of Fig. \ref{summary} $b)$ where only one electron remains unpaired and $S={1\over2}$.  The ground state is a doublet labelled as $^{2}E$ in Fig. 1 $a)$.

The energy scheme as a function of the energy splitting $10Dq$ between the $|t\rangle$ and $|e\rangle$ states is also dependent on the internal intra-orbital Coulomb interaction between the electrons which can be cast in terms of the quantity $J(dd)$ outlined above.  Detailed calculations have been made by Tanabe and Sugano (Refs. \onlinecite{Tanabe54:9,Tanabe54:9_2}) and by Liehr (Ref. \onlinecite{Liehr63:67}). 

The calculations we have done are shown in Fig. \ref{summary} $a$) where we have calculated the Co$^{2+}$ energy levels as a function of crystal field strength ($10Dq$) taking the strong crystal field $|t\rangle$ and $|e\rangle$ states as our basis.  The calculations used the electrostatic matrices and character tables supplied in Ref. \onlinecite{Griffiths:book,McClure59:9}.  The calculations depend on the Racah parameters $B$ and $C$ and we find that values close to those provided by Ref. \onlinecite{Abragam:book,McClure59:9,Griffiths:book} provide a reasonable description of the data.

The energy eigenvalues calculated from a strong crystal field basis link up with the free ion values for small crystal fields as illustrated in Fig. \ref{summary} $a)$.  The calculation displayed in Fig. 1 also provides an estimate for the crystal field energy where the ground state switches from the high spin $^{4}T_{1}$ (predicted based upon a weak crystal field scheme taking the free ion states as the basis functions) to a low spin $^{2}E$ state (expected based upon a strong crystal field basis).   

The Tanabe-Sugano diagram is traditionally drawn as a function of $10Dq/B$, where $B$ is the Racah parameter.  This is because the ratio $C/B$ is close to a constant of $\sim$ 4.6 across the $3d$ transition metal ions.  To make the connection with the Hubbard model and recent LDA+U models, we have redrawn the Tanabe-Sugano diagram in terms of the quantity $10Dq/J(dd)$ while fixing the quantity $C(dd)$.   This analysis is justified by the fact that we have obtained three different energies allowing us to uniquely derive the three experimental quantities $10Dq$ (the crystal field splitting), $J(dd)$ (intra-orbital exchange), and also $C(dd)$ (the angular part) which can be compared with theory.   We can also work back (noting the definitions written above for the free ion solutions) to obtain the Racah parameters $B$ and $C$ which can be compared with works done on dilute and free ion systems and published in Refs. \onlinecite{Abragam:book,McClure59:9,Griffiths:book}.  This comparison is made later in the presentation and parameterization of the experimental results.

\subsection{Spin-orbit coupling}
 
A weaker energy scale is given by the spin-orbit coupling, represented by the following term in the Hamiltonian,

\begin{eqnarray}
H_{so}=\lambda \vec{L} \cdot \vec{S}.
\label{spin_orbit}
\end{eqnarray}

\noindent $\vec{L}$ and $\vec{S}$ are the orbital and spin angular momentum respectively.  $\lambda$ is the spin-orbit coupling constant.  For simplicity, we ignore here extra factors used to account for strong covalency.   Given that the $^{4}T_{1}$ ground state is separated by $\sim$ 1 eV from the first excited state, we consider only the effect of this Hamiltonian on the ground orbital triplet within a fixed manifold of constant $\vec{L}$ and $\vec{S}$.  

\begin{figure}[t]
\includegraphics[width=8.5cm] {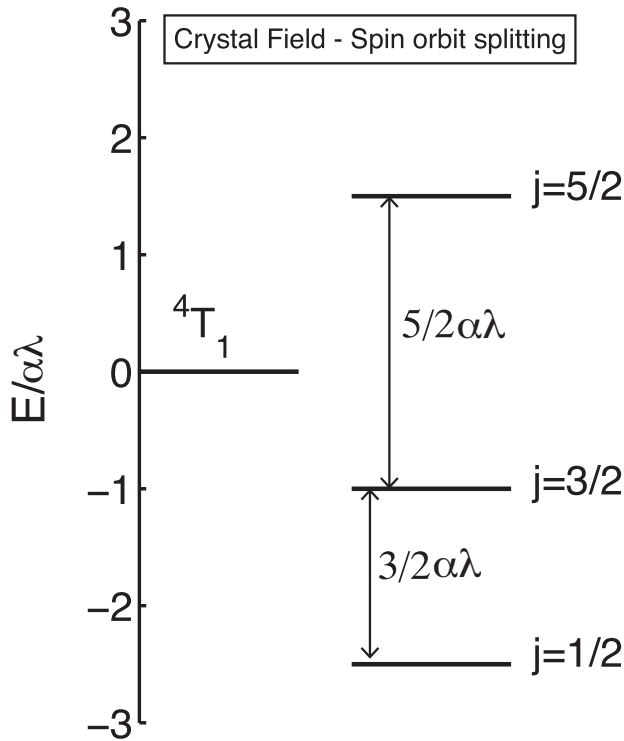}
\caption{\label{spin_orbit_diagram} The effect of the spin orbit term on the 12 fold degenerate (4 spin states $\times$ 3 orbital) $^{4}T_{1}$ ground state of CoO.  Assigning a fictitious orbital angular momentum of $l=1$ results in the crystal field scheme on the right.  The degeneracy of each spin-orbit split level can further be broken by the presence of a molecular field from neighboring ordered moments.  The experiment overcame this complication by studying the spin-orbit transitions in dilute samples of MgO doped with 3\% Co$^{2+}$.}
\end{figure}

For the orbital triplet ground state, it is useful to consider a total angular momentum $\vec{j}=\vec{l}+\vec{S}$ with an orbital angular moment of $l=1$ and orbital moment projection factor $\alpha=-{3/2}$.  The spin-orbit hamiltonian then becomes, 

\begin{eqnarray}
H_{so}=\alpha\lambda \vec{l} \cdot \vec{S}.
\label{spin_orbit}
\end{eqnarray}

\noindent With fixed magnitudes $|\vec{l}|\equiv l$ and $|\vec{S}| \equiv S$, the energy eigenvalues are then given by

\begin{eqnarray}
E={\alpha\lambda\over2}\left[j(j+1)-S(S+1)-l(l+1)\right].
\label{energies}
\end{eqnarray}

\noindent For $l=1$ and $S={3\over2}$ there are three energy eigenvalues illustrated in Fig. \ref{spin_orbit_diagram}.  The ground state corresponds to a doublet with $j= {1\over2}$ and with excited states of $j= {3\over2}$ and  ${5\over2}$.   The energy difference between the ground state and the lowest excited state is ${3\over2} \lambda \alpha$.  The selection rules for dipolar neutron scattering show that for the allowed transitions, the angular $j$ changes by 0 or $\pm$ 1.  This implies that at low temperature when only the lowest energy state is occupied, the spectra will show only transitions within the $j={1\over2}$ doublet and from those states to the $j={3\over2}$ states at an energy ${3\over2} \alpha\lambda$. 

The degeneracy of each level displayed in Fig. \ref{spin_orbit_diagram} can be split in the presence of a molecular field induced through magnetic order and superexchange.  The field on each Co$^{2+}$ site therefore depends on the magnetic structure and also the values of the nearest neighbor and next nearest neighbor superexchange values.  To avoid these extra terms we have measured a sample of MgO diluted with a small amount of Co$^{2+}$ to isolate and measure the pure spin-orbit term independent of the exchange parameters.  This is described in the following sections.

\section{Experimental Results}

This section discusses the experimental results and the connection with the crystal field theory outlined above.  As discussed earlier in the paper, the experiments were performed with the ISIS spallation neutron source that produces a burst of high-energy neutrons from a pulsed proton beam striking a tungsten target.  While the high-energy neutrons are moderated, a significant epithermal flux of neutrons above $\sim$ 1 eV remains.~\cite{Stock10:81,Kim11:84,Stock10:82}  High energy experiments where performed at the MAPS spectrometer to characterize the crystal field and intra-orbital exchange.  Lower energy experiments to determine the spin-orbit coupling were performed on the MARI and IRIS spectrometers. 

\subsection{Measured transitions and relation to Tanabe-Sugano Diagram and Racah Parameters}

The electronic ground state of the seven $3d$ electrons of a Co$^{2+}$ ion situated in CoO are largely determined by two energy scales - the intra-orbital exchange and the crystalline electric field.  The exchange can be characterized by a parameter $J(dd)$ where a large $J(dd)$ results in a ground state determined by Hund's rules.~\cite{Marel88:37}   The crystalline electric field is characterized by a parameter $10Dq$ which has the effect of splitting the  degeneracy of the $d$ orbitals.  

The energy eigenvalues as a function of  $10Dq/J(dd)$ is illustrated through the Tanabe-Sugano diagram (Fig. \ref{summary}) which displays the low-energy eigenvalues as a function of $10Dq/J(dd)$ for a Co$^{2+}$ in an octahedral environment.~\cite{Tanabe54:9, Tanabe54:9_2, Griffiths:book,McClure59:9}  The right side of the figure represents the strong crystal field limit where the doubly degenerate $e$ levels (at an energy of $6Dq$) and triply degenerate $t$ (at an energy of -$4Dq$) are taken as the basis set.  The critical field where the ground state changes occurs when the crystal field parameter $10Dq/J(dd) \sim$ 2.5 (shown by the vertical dashed line in Fig. \ref{summary}).   In small crystal fields Co$^{2+}$ adopts a high-spin ($^{4}T_{1},S={3\over2}$) configuration as given by Hund's rules while for large cubic crystal fields a low-spin configuration ($^{2}E, S={1\over2}$) is the ground state if the energy scale of the crystal field splitting dominates over the intra-orbital exchange $J(dd)$.

\begin{figure*}[t]
\includegraphics[width=15cm] {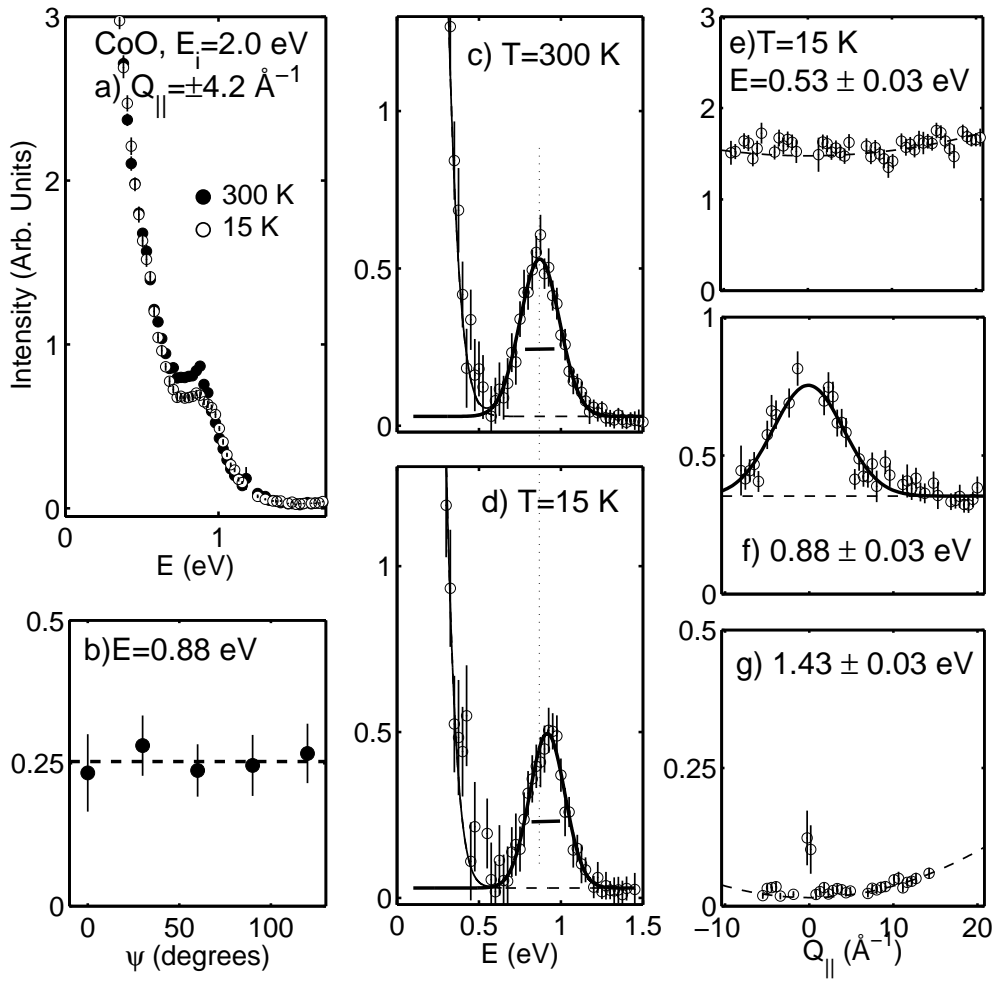}
\caption{\label{Ei_2eV} Neutron results at E$_{i}$=2.0 eV probing the high-energy single-ion crystal field excitations.  $a)$ Energy scans at 15 and 300 K integrating over the low- scattering angles.  $b)$ The angle dependence of the intensity of the 0.88 eV peak.  $c-d$) Background corrected constant Q scans taken at 300 K and 15 K.  $e-g)$ Momentum scans taken at several different energies.}
\end{figure*}

We first consider the high-energy electronic levels  and later study the much lower spin-orbit energy scale.  Figure \ref{Ei_2eV} shows data taken using the MAPS spectrometer on a single crystal of CoO with E$_{i}$=2.0 eV.  Uncorrected data integrating over small angles is plotted in panel $a)$ and background corrected constant-Q scans are shown in panels $c)$ and $d)$ at 300 K and 15 K in the disordered cubic and spin-ordered tetragonal phases respectively.  The background and magnetic components of the scattering were extracted through a series of constant energy scans, examples of which are illustrated in $e-g)$.  Each constant energy scan was fitted to a Gaussian centered at Q$_{||}$=0 plus a term modelled as $A+BQ^{2}$ to account for multiple scattering resulting from the phonons (seen in panel $g)$.  The momentum dependence allows us to separate magnetic scattering from background inconsistent with any electronic origin.

Panels $c)$ and $d)$ show magnetic peaks at 0.875 $\pm$0.009 eV and 0.917 $\pm$0.008 eV at 300 and 15 K respectively. $b)$ illustrates a plot of the intensity of the 0.88 eV (T=300 K) peak as a function of angle between the [001] axis and the incident beam.  0$^{\circ}$ represents the case with [001] parallel to $\vec{ k}_{i}$ and at 90$^{\circ}$, a [110] axis is parallel to $\vec{k}_{i}$.  No change in the intensity or the energy is observed.

 \begin{figure}[t]
\includegraphics[width=9.2cm] {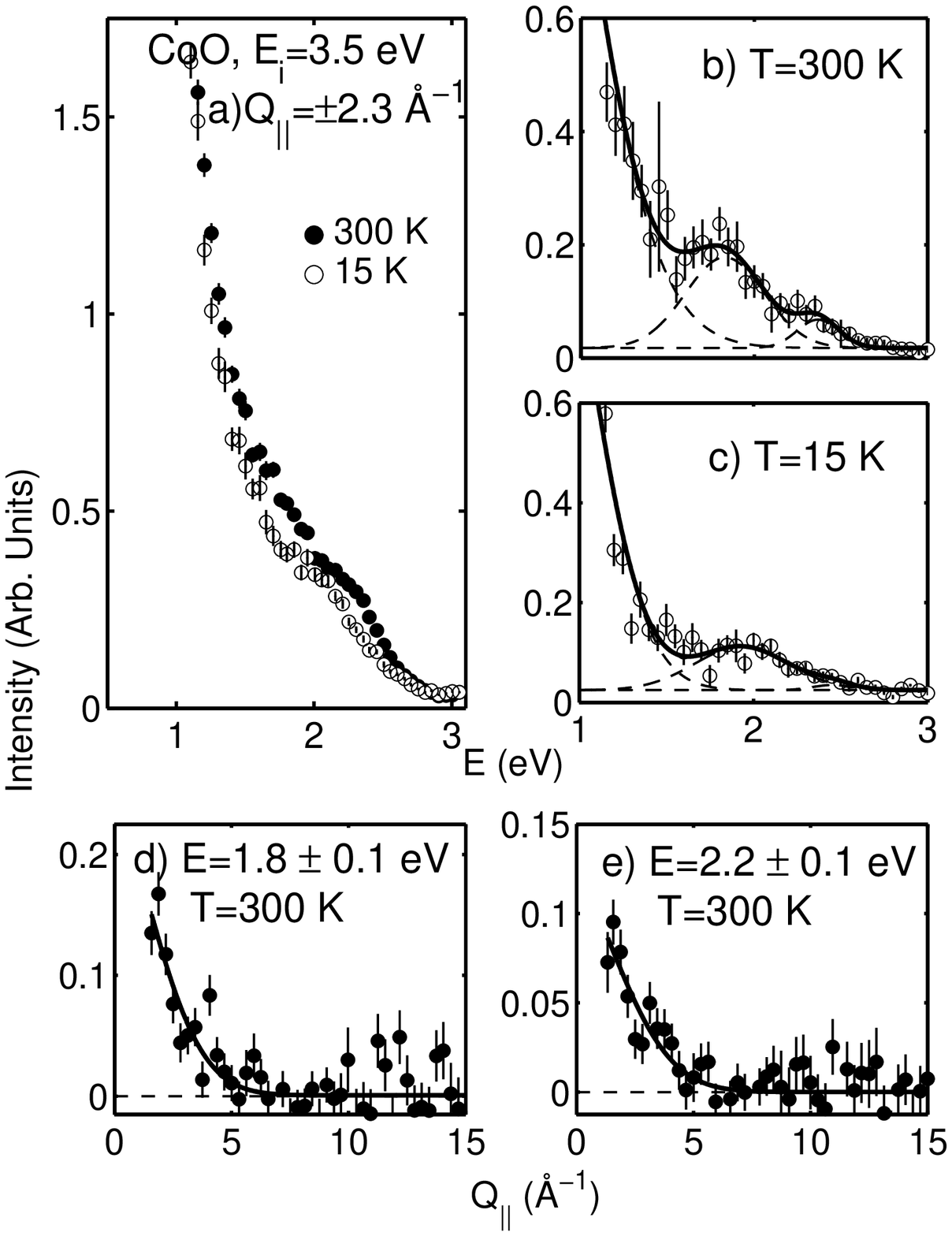}
\caption{\label{Ei_3p5eV}  $a)$ Constant Q scans at 15 and 300 K summing over the low-angle detector banks.  $b-c)$ illustrate background subtracted scans at 300 and 15 K. $d-e)$ plot constant energy scans at 300 K over the energies indicated.  The strong decay with momentum illustrates the magnetic nature of the scattering.  Note that the $x$ axis is defined in terms of one momentum transfer element $Q_{ll}$ as defined above.}
\end{figure}

Figure \ref{Ei_3p5eV} illustrates a similar series of scans taken with E$_{i}$=3.5 eV.  $a)$ shows the uncorrected data with $b)$ and $c)$ plotting background corrected scans in the cubic (300 K) and tetragonal (15 K) phases respectively. Panels $d)$ and $e)$ illustrate constant energy scans and show decreasing intensity with increasing momentum transfer. This proves the electronic origin of these excitations.  At 300 K, peaks occur at 1.84$\pm$0.03 and 2.3$\pm$0.1 eV.  In the ordered tetragonal phase at 15 K, a shift in the energy spectrum is measured with a broad peak observed at 1.93 $\pm$ 0.04 eV (panel $c$).  

We shall initially describe the results for the neutron scattering at high temperatures in the cubic phase (300 K). Three peaks were observed with energies below 3 eV namely; 0.870$\pm$0.009 eV, 1.84$\pm$0.03 eV, and 2.3$\pm$0.15 eV.  A comparison with calculations is shown in Fig. \ref{summary}.  The diagram requires the input of 3 parameters. Two parameters are the Racah parameters $B,C$ or Hubbard parameters $J(dd),C(dd)$ and describe the electronic structure in the absence of a crystal field term.  The third parameter is the crystal field term ($10Dq$) and describes the splitting between the $|e\rangle$ and $|t\rangle$ levels (Fig. \ref{summary}$b)$).    The diagram does not depend on the large Hubbard $U$ and our experiments are not sensitive to this value.  

A result of crystal field theoretical and experimental work is that the ratio of ${C\over B}$ and also ${J(dd)\over C(dd)}$ are roughly a constant for $3d$ transition metal ions and therefore the diagram is qualitatively fixed through just two parameters - $10Dq$ (crystal field splitting) and $J(dd)$ (intra-orbital exchange).~\cite{McClure59:9,Marel88:37}  Our experiment is compared with the calculation by varying these two parameters by the shaded areas that show the energies of the excitations while the size of the boxes are representative of the errors in the measured energies and the fitted values of $10Dq/J(dd)$.  The data is inconsistent with the energies on the right hand side of the figure and hence show that the ground state is the high-spin $^{4}T_{1}$ triplet, as expected from the large spontaneous magnetization of the antiferromagnetic state.   

The energy scheme is reproduced by the calculation if,

\begin{eqnarray}
J(dd)=1.3\pm0.2 eV \\
C(dd)= 0.49\pm\nonumber 0.10 eV\\
{C(dd)\over J(dd)}=0.40\pm0.15 \nonumber,
\label{Hubbard}
\end{eqnarray}

\noindent and the crystal field splitting is

\begin{eqnarray}
10Dq=0.94\pm0.10 eV. \nonumber
\label{Racah}
\end{eqnarray}

\noindent The value of $J(dd)$ agrees well with the local density approximation (LDA) calculations that reported 0.92 eV (\onlinecite{Anisimov91:44}) and the interpolation formula which give a value of $\sim$ 1.0 eV (Ref. \onlinecite{Marel88:37}).  The third parameter was also allowed to vary and is in good agreement with theory as ${C(dd) \over J(dd)}=0.40 \pm 0.15$ comparing well with the calculated values of $\sim$ 0.5.~\cite{Marel88:37}  Again, our experiment is not sensitive to the Hubbard $U$ estimated at $\sim$ 2 eV.   The parameters also compare very well with calculations presented in Ref. \onlinecite{Anisimov91:44} which suggest a value of $J(dd)$=0.92 eV and also the Hartree-Fock result - ($0.81+0.08 (Z-21)=1.29$ eV) (where $Z$ is the atomic number) and empirical relations ($0.59+0.075 (Z-21)=1.04$ eV) stated in Ref. \onlinecite{Marel88:37}.

Based on our high energy experiments, we find that CoO can best be described in the intermediate to weak crystal field limit in which $^{4}T_{1}$ is the ground state.  CoO is therefore far from a high-low spin transition. However, we note that while the crystal field parameters are in good agreement with theory, the crystal field gap $10Dq$ is roughly the same size as $J(dd)$.  Therefore, while CoO maybe considered in the weak crystal field limit, the crystal field energy is sizable and comparable to the intra-orbital exchange. 

Using the equations outlined above regarding the relevant crystal field theory, we can also write the parameters in terms of Racah parameters. We therefore derive the following values for Co$^{2+}$ in CoO from the parameterization listed above,

\begin{eqnarray}
B=0.11\pm0.02 eV \nonumber \\
C=0.62\pm0.10 eV \nonumber \\
{C \over B}=5.6\pm 1.1 \nonumber
\label{Racah}
\end{eqnarray}
 
\noindent This should be compared with $B$=0.13 eV and $C$=0.64 eV quoted in the summary listed in Ref. \onlinecite{McClure59:9}.  The ratio of ${C \over B}=5.6$ is close to the values derived for Co$^{2+}$ in molecules which is typically around 4.8.~\cite{Abragam:book,McClure59:9}  

\subsection{Neutron Intensities for $dd$ excitations in the dipolar approximation}

In the previous section, it was discussed how we calculated the energy eigenvalues in terms of a modified Sugano-Tanabe diagram shown in Fig. \ref{summary}.  Our experiment observed energies at 0.870$\pm$0.009 eV, 1.84$\pm$0.03, and 2.30$\pm$0.15 eV.  Based on these energies we derived crystal field parameters that are in good agreement with previous experiment and theory.  Furthermore, the 0.87 eV and 2.3 eV excitations are in excellent agreement with non-resonant inelastic x-ray experiments.~\cite{Larson07:99,Haverkort07:99}  

We now discuss the neutron intensities and selection rules for the transitions measured above and expected based upon the analysis in Fig. \ref{summary}.  Transitions based upon matrix elements of the form $\langle 0|M_{\pm,z}|f\rangle$ (expected in the dipolar approximation) were calculated so as to determine which transitions have a finite cross section.  Note that $\vec{M}=\vec{L}+2\vec{S}$. The eigenstates of the initial and final states were determined by relying on the tables provided in Refs. \onlinecite{Abragam:book,Griffiths:book}.   These give the eigenstates of the levels in question (Figs. \ref{summary}, \ref{Ei_2eV}, and \ref{Ei_3p5eV}) in terms of Slater determinants of the basis states $|l=2, m \rangle$.  This corresponds to calculating the matrix elements from both orbital and spin operations of $L_{z,\pm}$ and $S_{z,\pm}$.  

An orbital transition to the excited $^{4}T_{2}$ level is allowed for dipolar transitions and accounts for the 0.870$\pm$0.009 eV excitation (shown in Fig. \ref{Ei_2eV}).    Based on its energy, the 1.84$\pm$0.03 eV excitation originates from either the $^{2}E$ or $^{4}A_{2}$ level, but both are forbidden for dipolar matrix elements.  However, a transition involving a large quadrupolar matrix element is allowed to the $^{4}A_{2}$ and has been predicted but not resolved with x-ray experiments.~\cite{Haverkort07:99}  The 2.3$\pm$0.1 eV peak corresponds to a $^{4}T_{1}$ excitation which has been observed with x-rays previously and also calculated to have a finite quadrupolar matrix element.  

While this analysis is for the cubic phase, below 300 K CoO is antiferromagnetically ordered and tetragonal and it is expected that the  magnetic molecular field will lower the ground state energy. Figures \ref{Ei_2eV} and \ref{Ei_3p5eV} show that the two lowest energy excitations increase on cooling by 0.047$\pm$0.014 eV and 0.09$\pm$0.05 eV, respectively.  Although neither the magnetic molecular field nor the tetragonal distortion are known, the increases in the energy of the excitations are similar to the changes in energy expected from magnetic ordering given the estimated exchange constants.~\cite{Sakurai68:167,Kanamori59:87}

\subsection{Direct measure of spin-orbit coupling in dilute samples of MgO}

Having presented the high energy crystal field scheme characterizing the crystal field component of the Hamiltonian, we now discuss the second important energy scale referred to above, namely the spin-orbit coupling; $H=\lambda \vec{L}\cdot\vec{S}$.  As outlined above, CoO is cubic in the paramagnetic phase, and the orbital triplet ground state of the Co$^{2+}$ ion can be represented by a fictitious $l$=1 with  $\vec{L} =  \alpha \vec{l}$, where $\alpha=-{3 \over 2}$. Within the manifold of the $^{4}T_{1}$ states, the spin-orbit coupling becomes $H= -{3 \over 2} \lambda \vec{l}\cdot\vec{S}$ and splits the 12 states into 2 degenerate states with $j ={1 \over 2}$, 4 with $j = {3 \over 2}$ and 6 with $j = {5 \over 2}$ where $\vec{j} = \vec{l} +\vec{S}$. The energy difference between the ground state with $j = {1 \over 2}$ and the excited state with $j ={3 \over 2}$ is $-\lambda {9 \over 4}$ (note that $\lambda$ is negative).  These energy levels cannot be directly measured in pure CoO because of large magnetic exchange terms which split the degeneracy of the energy levels summarized in Fig. \ref{spin_orbit_diagram} therefore greatly complicating the excitation spectrum measured with neutrons. We have overcome this by studying dilute samples of Mg$_{0.97}$Co$_{0.03}$O for which most of the Co$^{2+}$ atoms are isolated from another Co atom and because MgO remains cubic down to the lowest temperatures measured (5 K) therefore removing the need to include terms due to the crystal field distortion.

 \begin{figure}[t]
\includegraphics[width=9.2cm] {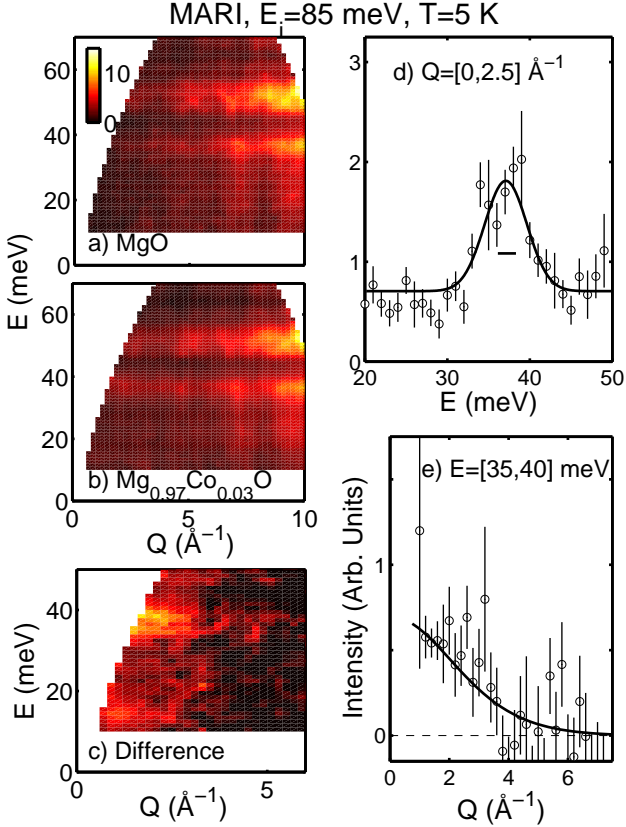}
\caption{\label{spin_orbit}  The inelastic response in MgO and dilute Mg$_{0.97}$Co$_{0.03}$O at 5 K. $a)$ and $b)$ illustrates the response in pure MgO and dilute Mg$_{0.97}$Co$_{0.03}$O respectively.  $c)$ shows a subtraction illustrating the response from dilute Co$^{2+}$.  $d)$ and $e)$ plot Q and energy averaged cuts demonstrating a peak at 37.1 $\pm$ 0.5 meV and that it decays with momentum transfer as expected from the form factor for the Co$^{2+}$ magnetic moment (solid line).  The magnetic peak positions was used to extract the spin-orbit coupling constant for CoO.  The use of dilute samples avoids complications due to the exchange terms in the Hamiltonian.}
\end{figure}

Fig. \ref{spin_orbit} shows high resolution results for powders of MgO and  Mg$_{0.97}$Co$_{0.03}$O with $a)$ and $b)$ plotting energy and momentum slices. The MgO spectrum is dominated by two phonon bands at $\sim$ 40 and 50 meV, however the doped samples also shows additional scattering at small momentum transfers illustrated by the subtraction in Fig. \ref{spin_orbit} $c)$.  Fig. \ref{spin_orbit} $d)$ and $e)$ show constant wave vector cuts and constant energy scans through the band of excitations with a peak at $E_{0}$=37.1$\pm$0.5 meV and with a full width of 6.1$\pm$1.0 meV. The energy width is considerably broader than the resolution width (1.8 meV) due to the different environments around each Co$^{2+}$ ion. The momentum dependence ($e)$ shows a decrease in the intensity confirming the magnetic origin of this intensity where the solid line is the square of the Co$^{2+}$ form factor.~\cite{Brown:tables}  The energy difference between the $j={1\over2}$ and $j={3\over2}$ levels, $E_{0}$ then fixes the spin-orbit coupling as $\lambda= -{4\over9}E_{0}$= -0.016$\pm$ 0.003 eV. The intensity of the transition from the $j ={1\over2}$ level to the $j = {5\over2}$ is forbidden for the neutron scattering cross section, while the transition from $j={3\over2}$ to $j={5\over2}$ is suppressed at low temperatures. 

The theoretical free-ion value for $\lambda$ is -0.0234 eV while the experimental value is listed as -0.0221 eV.~\cite{Abragam:book} Indirect and model dependent attempts at extracting $\lambda$ in nominally pure CoO have been made and give varied results. Soft x-ray and infrared techniques (Ref. \onlinecite{Schooneveld12:116,Kant08:78}) give -18.9 meV and a spin-wave analysis based upon the lower magnetic branches gave -12 meV (Ref. \onlinecite{Tomiyasu04:70}).   These results need to include the values of the exchange parameters which requires a detailed knowledge of the momentum dependence of the spin excitations and the various exchange constants along with terms accounting for the structural distortion at low temperatures.  All results, including here, show systematically smaller values than the free ion values.   This decrease could be indicative of coupling to the excited $^{4}P$ state, Fig. \ref{summary} or be due to covalency with oxygen ions.  The lack of any new excitations at high-energies, such as charge transfer excitations involving the O$^{2-}$ ions, is suggestive that the reduced spin-orbit coupling is the result of mixing between the ground state and the excited  $^{4}P$ levels.  The role of charge-transfer excitations and the relevant cross section for neutrons requires further investigation.

\subsection{Search for low-energy crystal fields}

While the energy scale of the orbital excitations is expected to be large, some neutron scattering results (Ref. \onlinecite{Feygenson11:83}) have suggested the presence of very low-energy peaks of the order $\sim$ 1 meV and crystal field excitations with these low energies are inconsistent with the crystal field description provided above along with the comparison with LDA+U models.  We have performed a search for these excitations using both direct and indirect geometry neutron spectrometers and discuss the results here.

\begin{figure}[t]
\includegraphics[width=9.5cm] {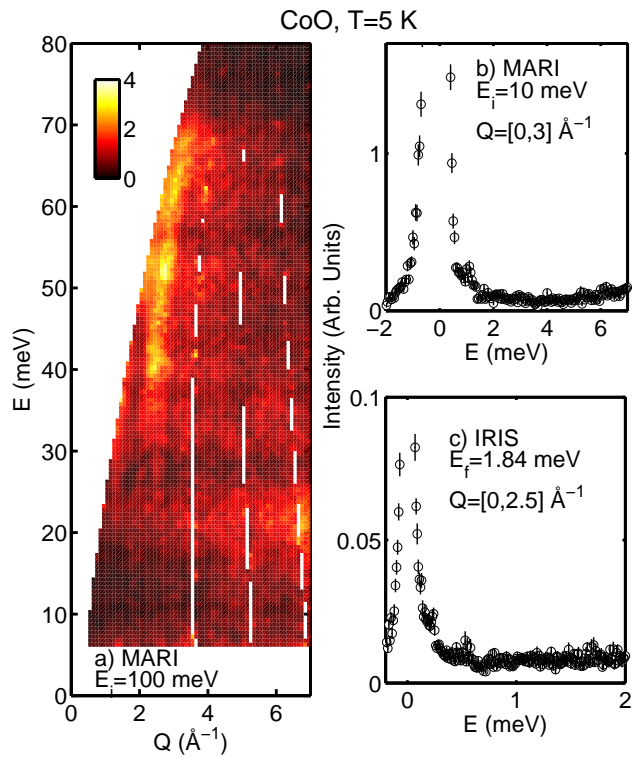}
\caption{\label{MARI_IRIS} $a)$ Powder averaged $Q-E$ map of the low-energy (E$<$100 meV) scattering from a powder of CoO.  The one-magnon cross section is readily observed between $\sim$ 30 meV and 70 meV.  $b)$ and $c)$ illustrate the absence of any obvious extra peak at low energies below $\sim$ 6 meV.}
\end{figure}

To search for possible low-energy crystal field excitations, we performed a series of experiments using both powder and single crystals of CoO.  Figure \ref{MARI_IRIS} illustrates the scattering from a powder averaged $Q-E$ map of CoO measured with the MARI spectrometer.  There is a band of excitations from $\sim$ 30-70 meV, previously investigated in other papers (Refs. \onlinecite{Yamani08:403,Yamani10:88,Tomiyasu06:75}), and associated with the spin/orbital excitations.  A weak phonon band, also previously measured for single crystals (Ref. \onlinecite{Sakurai68:167}), is seen at $\sim$ 20 meV.  Below this band, there is little evidence for magnetic scattering as demonstrated in panels $b)$ and $c)$.  While some low-energy peaks ($\sim$ 1-2 meV) were visible, we found these peaks changed with the incident energy and experimental configuration and therefore associate them with double scattering from shielding walls around the sample position.  The peaks are also sharper than both the measured and calculated experimental resolutions in energy.   We therefore have no evidence for the low-energy spin excitations claimed to be seen in Ref. \onlinecite{Feygenson11:83} below the gapped ($\sim$25 meV) one-magnon excitations described previously.~\cite{Yamani08:403,Yamani10:88,Tomiyasu06:75,Sakurai68:167}

\begin{figure}[t]
\includegraphics[width=7.0cm] {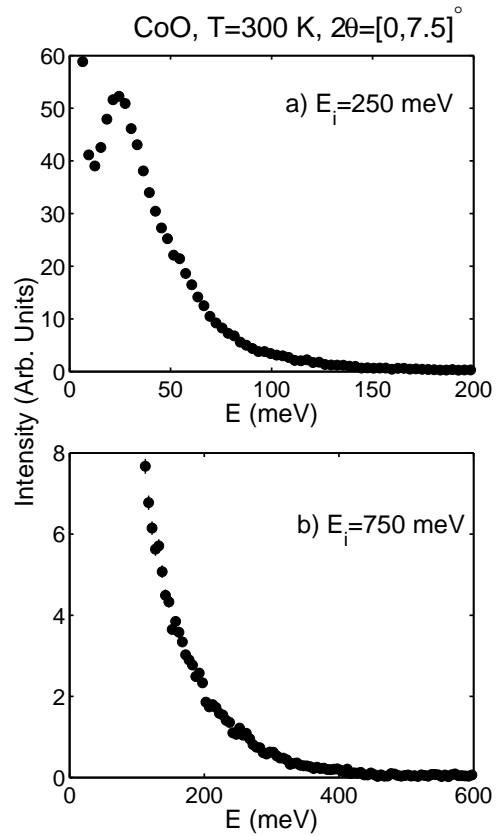}
\caption{\label{lowE} Scans performed on MAPS using a single crystal sample of CoO with $\vec{k}_{i}$ aligned along [001].  $a)$ illustrates data taken with E$_{i}$=250 meV showing the spin/orbital excitations between $\sim$ 20-75 meV. $b)$ shows a similar scan taken with E$_{i}$=750 meV demonstrating no obvious crystal field peaks for energy transfers between 100 and 600 meV.}
\end{figure}

For completeness, we continued the search for crystal field excitations, above the top of the one-magnon band of $\sim$ 70 meV, using the MAPS chopper spectrometer.  Figure \ref{lowE} illustrates data taken with a single crystal of CoO using the MAPS spectrometer with $\vec{k}_{i}$ aligned along [001].  Fig. \ref{lowE} $a)$ illustrates data taken integrating over low angles and illustrates the presence of significant magnetic scattering in the range of $\sim$ 20-75 meV.  We associate this scattering with excitations within the $^{4}T_{1}$ ground state referred to as the one-magnon above and shown in Fig. \ref{MARI_IRIS} $a)$.  These have been discussed in detail in several other papers but are not discussed in detail here.~\cite{Yamani08:403,Yamani10:88,Tomiyasu06:75,Sakurai68:167}  Fig. \ref{lowE} $a)$ does illustrate the absence of any additional peaks between 100 and 200 meV.  A similar scan is illustrated in Fig. \ref{lowE} $b)$ and further indicates a sloping background with no obvious peaks below $\sim$ 600 meV.  Based upon these two scans performed on MAPS, we conclude there are no crystal field excitations between 100 and 600 meV.  Based on the low-energy searches performed on the MAPS, MARI, and IRIS spectrometers. We conclude there is no observable magnetic scattering below the gapped one-magnon scattering extending over the energy range from $\sim$20 to 70 meV.  This conclusion agrees with our crystal field analysis presented above and also with corroborating calculations based upon band theory.

\section{Discussion}

The previous section described the single ion excitations and derived the relevant parameters in terms of classic crystal field theory (Racah parameters and Tanabe-Sugano diagrams) and also in terms of more recently developed LDA+U band theories.  We also used dilute samples to extract the spin-orbit coupling.  The use of dilute samples avoided complications due to spin and orbital exchange which then makes it difficult to understand the neutron cross section and excitation spectrum.  We now summarize these results by comparing them with existing x-ray and optical techniques.

\subsection{Comparison with inelastic x-rays and optical techniques}

We now compare neutron and inelastic  x-ray results.  Non-resonant x ray experiments (Ref. \onlinecite{Larson07:99}) gave peaks in CoO at $\sim$ 1.0 and 2.3 eV and similar energies were derived from infrared and soft x-ray experiments (Ref. \onlinecite{Pratt59:116,Kant08:78,Schooneveld12:116}).  The excitations were measured and had significant anisotropy. They were understood in terms of band structure calculations requiring a large electronic gap $\sim$ 3 eV.~\cite{Haverkort07:99}  Measurements using other x-ray resonant spectroscopy techniques have found further peaks located just below $\sim$ 2.0 eV.

We observe three excitations, in the cubic phase, at 0.870$\pm$0.009 eV, 1.84$\pm$0.03, and 2.3$\pm$0.1 eV and find no angular dependence or dispersion within error.   The 0.870 and 2.3 eV peaks are in agreement with x-rays but the 1.84 eV excitation was not resolved with non-resonant x-rays and is forbidden in the dipolar approximation, however it has been reported in other optical and x-ray studies such as RIXS and EELS that are sensitive to dipole forbidden cross sections.~\cite{Schooneveld12:116,Fromme98:57,Fiebig03:93,Chiuz08:78, Magnuson02:65, Gorsch94:49}  This result highlights the point that the 1.84 $\pm$0.03 eV excitation is dipole forbidden and confirms our claim that we are sampling a quadrupolar matrix element in our neutron scattering experiments.

One difference between the neutron results and the inelastic x-ray results is the range of momentum probed.  Due to kinematic constraints, our experiments occur at larger momentum transfers (approximately a factor 2) than those reported using non-resonant inelastic x-rays ($Q\sim$ 13 \AA$^{-1}$ here compared with 3.5 \AA$^{-1}$ with x-rays).  In this region, as pointed out in Ref. \onlinecite{Haverkort07:99}, other terms in the cross section need to be considered beyond the dipolar term and in particular the quadrupolar matrix element.  While there is no allowed excitation in the dipolar approximation, quadrupolar terms do allow a finite intensity for the $^{4}A_{2}$ level.  A similar result is predicted for the 2.3$\pm$0.1 eV which corresponds to a $^{4}T_{1}$ level.  We note that a neutron study of NiO required only the use of dipole terms to understand the spectrum, however, only the lowest energy terms were probed.~\cite{Kim11:84}  The larger momentum transfers probed with neutrons here consequently imply that our experiments are sensitive to a different lengthscale to that of inelastic x-rays.  This may provide an explanation for the lack of angular dependence in the intensity.

\subsection{Conclusions and summary}

The data can be consistently interpreted in terms of crystal field excitations, in agreement with calculations based upon LDA+U as well as early works parameterized in terms of the Tanabe-Sugano formalism.  The results are orbital excitons within the Mott-Hubbard gap and do not probe the large U expected in initial x-ray results.   We do not observe evidence of new modes that would be expected for charge transfer excitations (as observed in Mott Insulating cuprates~\cite{Kim02:89,Hasan00:288}) or for propagating Zhang-Rice singlet states.  This suggests that these modes reside at different energies than were probed here or that they are weaker than our sensitivity.~\cite{Yin08:100}  However, our results show that CoO is well described in terms of the class of a large-$U$, Hubbard Mott insulators with crystal field excitons below the Mott Hubbard gap.  The observation of dipole forbidden transitions also demonstrates the use of neutrons in measuring such cross sections. 

We are grateful to the STFC, CIFAR, and the Leverhulme Foundation for support.


%

\end{document}